# Fifteen classes of solutions of the quantum two-state problem in terms of the confluent Heun function

A.M. Ishkhanyan and A.E. Grigoryan

Institute for Physical Research, NAS of Armenia, 0203 Ashtarak, Armenia

We derive 15 classes of time-dependent two-state models solvable in terms of the confluent Heun functions. These classes extend over all the known families of three- and two-parametric models solvable in terms of the hypergeometric and the confluent hypergeometric functions to more general four-parametric classes involving three-parametric detuning modulation functions. In the case of constant detuning the field configurations describe excitations of two-state quantum systems by symmetric or asymmetric pulses of controllable width and edge-steepness. The classes that provide constant detuning pulses of finite area are identified and the factors controlling the corresponding pulse shapes are discussed. The positions and the heights of the peaks are mostly defined by two of the three parameters of the detuning modulation function, while the pulse width is mainly controlled by the third one, the constant term. The classes suggest numerous symmetric and asymmetric chirped pulses and a variety of models with two crossings of the frequency resonance. We discuss the excitation of a two-level atom by a pulse of Lorentzian shape with a detuning providing one or two crossings of the resonance. We derive closed form solutions for particular curves in the 3D space of the involved parameters which compose the complete return spectrum of the considered two-state quantum system.

## 1. Introduction

A few-state description is a good approximation of a real quantum system involved in the interaction with radiation if a few of its quantum levels are resonant or nearly resonant with the driving field, while the remaining levels are far off resonance. The analytic solutions of the two-state problem [1-3] have played an important role in the study of a number of physical phenomena in many branches of contemporary physics within the few-state representations. Many such solutions have been explored in the past using the hypergeometric functions, confluent hypergeometric functions and other familiar special mathematical functions (see, e.g., [1-10] and references therein). Nowadays, the search for analytic solutions still deserves attention since the numerical simulations in some cases may be of insufficient generality because of the large number of parameters involved or specific implicit singularities.

In the present paper, we discuss the solutions of the two-level problem in terms of the confluent Heun function, a member of the Heun class of mathematical functions that are believed to compose the next generation of special functions [11-13]. This function is the solution of the confluent Heun equation which is of particular interest because it directly incorporates the Gauss hypergeometric and the Kummer confluent hypergeometric equations, as well as the algebraic form of the Mathieu equation [11,12]. Other known equations can be viewed as particular, transformed or limiting cases of this equation, e.g., the spheroidal, Coulomb spheroidal, generalized spheroidal wave equations, and the Whittaker-Hill equation [11-13]. For this reason, one may expect that the analytic models solvable in terms of the confluent Heun function will directly generalize many of the known solvable cases. We will see that, indeed, this is the case; for instance, the derived classes cover all the previously known two-state models solvable in terms of hypergeometric and confluent hypergeometric functions. In addition, we obtain several new classes of models not treated before.

A distinct feature of the confluent Heun equation useful for general classification of the solvable models is that the mentioned particular equations are obtained by simple choices of separate parameters, without limiting procedures involving different parameters. This allows one to readily follow the origin of a particular extension when passing from a known model to a more general Heun one. For instance, due to the structure of its singularities (two regular singular points located at finite points of the complex $z$-plane as it is in case of the Gauss hypergeometric equation, and an irregular singularity of rank 1 located at infinity, as it is in the case of the confluent hypergeometric equation) the confluent Heun equation suggests a clear way to identify qualitative features coming from different singularities when generalization of the hypergeometric models is discussed. Note that the behavior of the corresponding quantum systems described by the confluent Heun equation is in general expected to be of the form of concurrence, constructive or destructive, of different features separately originating from the two prototype hypergeometric equations.

To find the field configurations for which the governing time-dependent Schrödinger equations are reduced to the confluent Heun equation, we apply the approach of [14-16] where the transformation of the dependent variable is used to derive the basic models which afterwards generate different families of field configurations via application of the transformation of the independent variable. This leads to the generalization of all known families solvable in terms of simpler special functions to more general classes and yields several new families of models integrable in terms of irreducible confluent Heun functions.

In total, fifteen classes of solvable models are derived. For each of the classes, the actual field configurations are generated by a pair of functions, one of which (referred to as the amplitude modulation function) stands for the amplitude of the field and the other one (referred to as the detuning modulation function) defines the variation of the frequency detuning. Though the classes are identified by the amplitude modulation function only, since the detuning modulation function is of the same form for all the derived classes, many of the particular properties of the field configurations are due to the detuning modulation function. For instance, in the case of a *constant detuning* field configuration the detuning modulation function defines the appropriate transformation of the independent variable which then, in combination with the amplitude modulation function, generates the corresponding pulse shape. Alternatively, in the case of a *constant amplitude* field configuration, when the transformation of the independent variable is defined solely by the amplitude modulation function, the detuning modulation function provides the particular shape of the time variation of the field's frequency detuning.

A notable feature provided by the utilization of the confluent Heun functions is the generalization of the previously known one- and two-parametric detuning modulation functions to the three-parametric case. This turns out to be useful in several instances. For example, in the case of constant detuning this leads to two-peak symmetric or asymmetric pulses with controllable width. Among these, rectangular box pulses of given width and infinitely narrow pulses are possible as limiting cases. Furthermore, in the general case of variable detuning a variety of level-crossing models are derived, including symmetric and asymmetric chirped pulses with two time scales [17], models of non-linear sweeping through the resonance [18], level-glancing configurations [19], processes with two resonance-crossing time points [20] and, in specific cases, multiple (periodically repeated) crossing models [21].

The level-crossing models have been widely studied in the context of many physical and chemical phenomena (see, e.g., [1-10,17-36]). A useful generic feature of such models described by the confluent Heun equation is that, due to additional parameters involved in the Heun equation, they in general suggest processes with more time scales compared with the models described by the hypergeometric equations. For instance, in the case of chirped pulses we have a coupling that acts over a time interval which is not connected to the effective time of the resonance crossing. Another representative example is the case of double crossing models where the time separation between two crossings and the speed at which the system crosses a particular resonance point are controlled almost independently, by separate parameters. The same advantageous feature is observed when periodically repeated crossings are discussed: the coupling strength, the period of crossings and the detuning modulation amplitude are described by well identified separate parameters. For this reason, below we discuss the possible crossing models and explore the basic properties of the corresponding field configurations in terms of the parameters of the confluent Heun equation.

## 2. Fifteen basic models solvable in terms of the confluent Heun function

The semiclassical time-dependent two-state problem is written as a system of coupled first-order differential equations for probability amplitudes of the two states $a_{1,2}(t)$ containing two arbitrary real functions of time, $U(t)$ (the Rabi frequency, $U > 0$) and $\delta(t)$ (detuning modulation, the derivative of which $\delta_t = d\delta/dt$ is the frequency detuning):

$$ia_{1t} = Ue^{-i\delta}a_2, \quad ia_{2t} = Ue^{+i\delta}a_1. \tag{1}$$

Here and below the lowercase Latin index denotes differentiation with respect to corresponding variable. System (1) is equivalent to the following linear second order ordinary differential equation:

$$a_{2tt} + \left(-i\delta_t - \frac{U_t}{U}\right)a_{2t} + U^2 a_2 = 0. \tag{2}$$

It can be shown that if the function $a_2^*(z)$ is a solution of this equation rewritten for an auxiliary argument $z$ for some functions $U^*(z)$, $\delta^*(z)$ then the function $a_2(t) = a_2^*(z(t))$ is the solution of Eq. (2) for the field configuration defined as

$$U(t) = U^*(z)\frac{dz}{dt}, \quad \delta_t(t) = \delta_z^*(z)\frac{dz}{dt} \tag{3}$$

for arbitrary complex-valued function $z(t)$ [14-16]. The pair of functions $U^*(z)$ and $\delta^*(z)$ is referred to as a basic integrable model.

Transformation of independent variable $a_2 = \varphi(z) u(z)$ together with Eq. (3) reduces Eq. (2) to the following equation for the new dependent variable $u(z)$:

$$u_{zz} + \left(2\frac{\varphi_z}{\varphi} - i\delta_z^* - \frac{U_z^*}{U^*}\right)u_z + \left(\frac{\varphi_{zz}}{\varphi} + \left(-i\delta_z^* - \frac{U_z^*}{U^*}\right)\frac{\varphi_z}{\varphi} + U^{*2}\right)u = 0. \tag{4}$$

This equation is the confluent Heun equation [11]

$$u_{zz} + \left(\frac{\gamma}{z} + \frac{\delta}{z-1} + \varepsilon\right)u_z + \frac{\alpha z - q}{z(z-1)}u = 0, \tag{5}$$

when

$$\frac{\gamma}{z} + \frac{\delta}{z-1} + \varepsilon = 2\frac{\varphi_z}{\varphi} - i\delta_z^* - \frac{U_z^*}{U^*} \tag{6}$$

and

$$\frac{\alpha z - q}{z(z-1)} = \frac{\varphi_{zz}}{\varphi} + \left(-i\delta_z^* - \frac{U_z^*}{U^*}\right)\frac{\varphi_z}{\varphi} + U^{*2}. \tag{7}$$

Eqs. (6) and (7) compose an over-determined system of two nonlinear equations for three unknown functions, $U^*(z)$, $\delta^*(z)$ and $\varphi(z)$. The general solution of this system is not known. However, many particular solutions can be found starting from specific forms of the involved functions. In the cases of the hypergeometric and confluent hypergeometric equations this approach has led to the extension of all previously known solvable cases to more general classes and has allowed the generation of several new classes [14-16]. Here we follow the steps suggested in these references and show that the technique is efficient also in the case of the confluent Heun equation.

Searching for solutions of equations (6), (7) in the following form:

$$\varphi = e^{\alpha_0 z} z^{\alpha_1} (z-1)^{\alpha_2}, \tag{8}$$

$$U^* = U_0^* z^{k_1} (z-1)^{k_2}, \tag{9}$$

$$\delta_z^* = \delta_0 + \frac{\delta_1}{z} + \frac{\delta_2}{z-1}, \tag{10}$$

we multiply Eq. (7) by $z^2(z-1)^2$ and note that it follows from the obtained equation that for arbitrary $\delta_{0,1,2}$ the product $U_0^{*2} z^{2k_1+2} (z-1)^{2k_2+2}$ is a fourth-degree polynomial in $z$. Hence, $k_{1,2}$ are integers or half-integers obeying the inequalities $-1 \le k_{1,2} \cup k_1 + k_2 \le 0$. This leads to 15 cases of $\{k_1, k_2\}$ which are shown in Fig. 1 by points in 2D-space of parameters $k_{1,2}$.

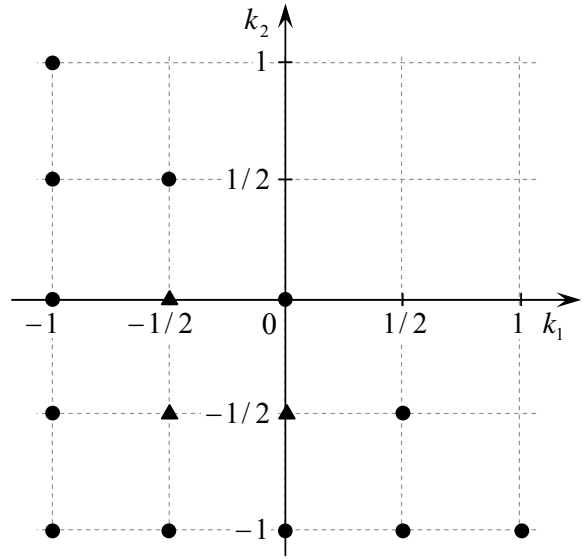

Fig. 1. Fifteen possible cases of $\{k_1, k_2\}$. The cases for which $\varphi(z) = 1$ are marked by triangles.

The corresponding basic models are explicitly presented in Table 1. We recall that the actual field configuration is given as

$$U(t) = U_0^* z^{k_1} (z-1)^{k_2} \frac{dz}{dt}, \tag{11}$$

$$\delta_t(t) = \left(\delta_0 + \frac{\delta_1}{z} + \frac{\delta_2}{z-1}\right)\frac{dz}{dt}. \tag{12}$$

Note that here the parameters $U_0^*$ and $\delta_{0,1,2}$ are complex constants which should be chosen so that the functions $U(t)$ and $\delta(t)$ are real for the chosen complex-valued $z(t)$. Since these parameters are arbitrary, all the derived classes are four-parametric in general.

Some of the obtained classes generate three-parametric subclasses of field configurations $\{U(t),\delta(t)\}$ for which the two-state problem is solvable in terms of hypergeometric or confluent hypergeometric functions. These classes are indicated in Table 1 by "$_2F_1$" and "$_1F_1$", respectively. Notably, two of the basic models, namely, $U^*/U_0^* = 1/z$ and $U^*/U_0^* = 1/(z-1)$, generate both a three-parametric subclass solvable in terms of $_1F_1$ (for that one should take $\delta_2 = 0$) and a three-parametric subclass solvable in terms of $_2F_1$ (in this case should be that $\delta_0 = 0$). Some other basic models allow *two*-parametric subclasses solvable in terms of hypergeometric or confluent hypergeometric functions, see below.

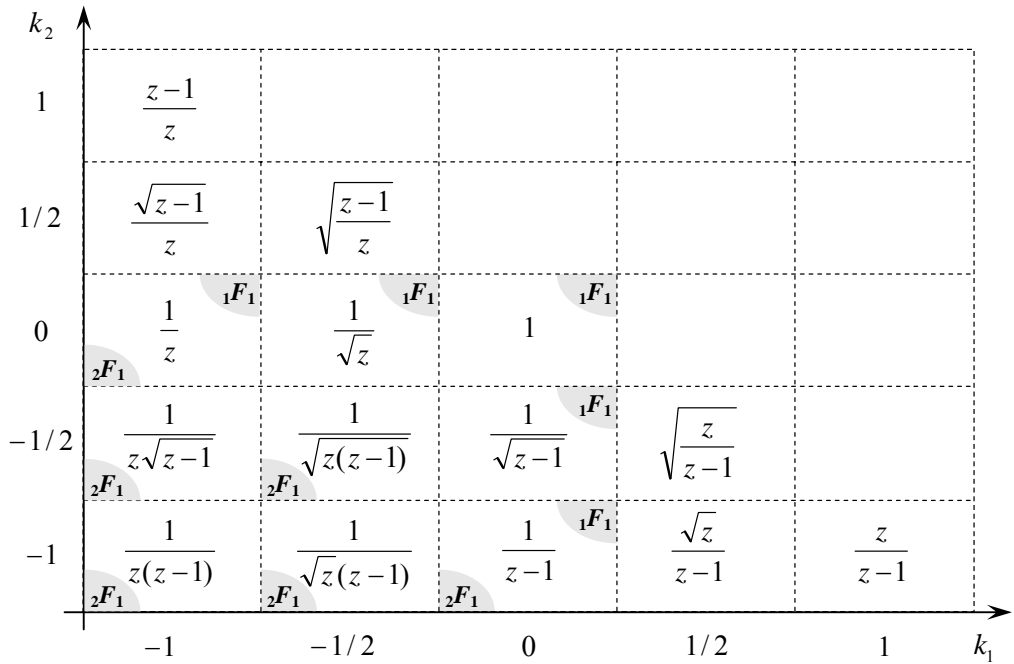

Table 1. Fifteen basic models of amplitude modulation function $U^*$ for which the two-state problem is solved in terms of the confluent Heun functions. The models that include three-parametric subclasses solvable in terms of hypergeometric and confluent hypergeometric functions are indicated by "$_2F_1$" and "$_1F_1$", respectively.

The basic models allowing three-parametric subclasses for which the two-state problem is solvable in terms of the Kummer confluent hypergeometric functions $_1F_1$ are $U^*/U_0^* = 1/z, 1/\sqrt{z}, 1, 1/\sqrt{z-1}$, and $1/(z-1)$ [14]. These families of field configurations correspond to the choice $\delta_1 = 0$ or $\delta_2 = 0$ in Eq. (12). Three-parametric subclasses of the classes $U^*/U_0^* = 1/\sqrt{z}$ and $1/\sqrt{z-1}$ specified by the choice $\delta_0 = 0$, $\delta_{1,2} \neq 0$, the solutions for which are not reduced to the hypergeometric functions, were recently presented in [37,38].

The six models in the lower left corner of Table 1, namely $U^*/U_0^* = 1/z$, $U^*/U_0^* = 1/(z\sqrt{z-1})$, $1/(z(z-1))$, $1/(\sqrt{z}(z-1))$, $1/(z-1)$, and $1/\sqrt{z(z-1)}$ include three-parametric subclasses of field configurations that allow solution in terms of the Gauss hypergeometric function $_2F_1$ (see [15,16]). These families correspond to the choice $\delta_0 = 0$ in the formula for $\delta_z^*$. It was shown that there exists a two-parametric subclass of the class $U^*/U_0^* = 1/(\sqrt{z}(z-1))$ with non-zero $\delta_0$: $\delta_0 + \delta_1 = -\delta_2/2$, $1 + \delta_2^2 = -4U_0^{*2}$, for which the solution is written in terms of the Kummer confluent hypergeometric function [14]. Because of the symmetry of the confluent Heun equation with respect to the interchange $z \leftrightarrow z-1$, a similar subclass can also be constructed for the class $U^*/U_0^* = 1/(z\sqrt{z-1})$.

Among the remaining six models $U^*/U_0^* = (z-1)/z$, $\sqrt{z-1}/z$, $\sqrt{(z-1)/z}$, $\sqrt{z/(z-1)}$, $\sqrt{z}/(z-1)$, $z/(z-1)$, two classes, $U^*/U_0^* = \sqrt{z}/(z-1)$ and $\sqrt{z-1}/z$, have two-parametric subclasses allowing solution in terms of the Kummer confluent hypergeometric functions [14]. For the first of these subclasses the specification of the parameters is $\delta_0 + \delta_1 = +\delta_2/2$, $1 + \delta_2^2 = -4U_0^{*2}$ [14]. Another two classes, $U^*/U_0^* = \sqrt{z/(z-1)}$ and $\sqrt{(z-1)/z}$ allow two-parametric subclasses, the solution for which is written in terms of the Gauss hypergeometric functions [15]. For the first of these subclasses the corresponding specification of the parameters is $\delta_0 = \pm 2U_0^*$, $\delta_2 = \delta_1 - \delta_0/2$ [15]. Thus, the only classes for which hypergeometric subclasses are not reported are $U^*/U_0^* = z/(z-1)$ and $U^*/U_0^* = (z-1)/z$.

The solution of the initial two-state problem is explicitly written as

$$a_2 = e^{\alpha_0 z} z^{\alpha_1} (z-1)^{\alpha_2} H_C(\gamma, \delta, \varepsilon; \alpha, q; z), \qquad (13)$$

where the confluent Heun function's parameters $\gamma$, $\delta$, $\varepsilon$, $\alpha$, $q$ are given as

$$\gamma = 2\alpha_1 - i\delta_1 - k_1, \ \delta = 2\alpha_2 - i\delta_2 - k_2, \ \varepsilon = 2\alpha_0 - i\delta_0, \qquad (14)$$

$$\alpha = -i\delta_0(\alpha_1 + \alpha_2 - \alpha_0) + \alpha_0(\gamma + \delta - \varepsilon) + Q'''(0)/6, \qquad (15)$$

$$q = \alpha_0(\alpha_0 - i\delta_0 - k_1 - i\delta_1) + \alpha_2(1 - \alpha_2 + k_1 + i\delta_1 + k_2 + i\delta_2) +$$

$$\alpha_1(1-\gamma-\delta+\varepsilon+\alpha_1)-Q''(0)/2-Q'''(0)/6 \tag{16}$$

with
$$Q(z) = U_0^{*2} z^{2k_1+2}(z-1)^{2k_2+2}$$

and
$$\alpha_0^2 - i\alpha_0\delta_0 = -Q^{(4)}(1)/4!, \tag{17.1}$$

$$\alpha_1^2 - \alpha_1(1+k_1+i\delta_1) = -Q(0), \tag{17.2}$$

$$\alpha_2^2 - \alpha_2(1+k_2+i\delta_2) = -Q(1). \tag{17.3}$$

## 3. Series solutions of the confluent Heun equation

A power series expansion of the solution of the confluent Heun equation:

$$H_C(\gamma,\delta,\varepsilon;\alpha,q;z) = z^\mu \sum_n c_n z^n, \tag{18}$$

is constructed using the following three-term recurrence relation for the coefficients [11-13]:

$$R_n c_n + Q_{n-1} c_{n-1} + P_{n-2} c_{n-2} = 0 \tag{19}$$

where
$$R_n = (\mu+n)(\mu+n-1+\gamma), \tag{20}$$

$$Q_n = q - (\mu+n)(\mu+n-1+\gamma+\delta-\varepsilon), \tag{21}$$

$$P_n = -\varepsilon(\mu+n) - \alpha. \tag{22}$$

For left-hand side termination of this series at $n=0$ one should put $c_{-1} = c_{-2} = 0$ and $R_0 = 0$, so that $\mu = 0$ or $\mu = 1-\gamma$. In some cases, the series is right-hand side terminated at some $n = N$. For this to occur, it is necessary to have $P_N = 0$, i.e., $\varepsilon(\mu+N)+\alpha = 0$, and $c_{N+1} = 0$. The latter equation is a polynomial equation of the order $N+1$ for the accessory parameter $q$ with, in general, $N+1$ solutions. Note that, in general, the convergence radius of the series is equal to unity.

Instead of powers, one may use other expansion functions to construct series solutions to the confluent Heun equation. Expansions in terms of descending powers as well as in terms of hypergeometric and confluent hypergeometric functions are well known [11-13,39-40]. Other examples include expansions in terms of the Hankel and Bessel functions [41], Coulomb wavefunctions [42], incomplete Beta functions [43]. Using the properties of the derivatives of the solutions of the Heun equation [44], expansions in terms of higher transcendental functions [45], e.g., the Goursat and the Appell generalized hypergeometric functions, can be constructed [46,47]. The expansions apply to different combinations of the involved parameters and have different regions of convergence. We mention here two particular expansions in terms of the confluent hypergeometric functions, which are convenient for derivation of closed form solutions applicable to the discussed two-state problem (see examples below) [39,40]. An expansion in terms of the Kummer confluent hypergeometric functions is [40]

$$H_C(\gamma,\delta,\varepsilon;\alpha,q;z) = \sum_{n=0}^{\infty} c_n \,_1F_1(\beta_0+n;\gamma;-\varepsilon z), \quad (23)$$

where the coefficients of the expansion are given by the recurrence relation

$$R_n c_n + Q_{n-1} c_{n-1} + P_{n-2} c_{n-2} = 0 \quad (24)$$

with

$$R_n = (n+\beta_0-\gamma)(n+\beta_0-\alpha/\varepsilon), \quad (25)$$

$$Q_n = (\gamma - 2(n+\beta_0))(n+\beta_0-\alpha/\varepsilon) + (n+\beta_0)(\varepsilon-\delta) - q, \quad (26)$$

$$P_n = (n+\beta_0)(n+\beta_0+\delta-\alpha/\varepsilon), \quad (27)$$

where $\beta_0 = \alpha/\varepsilon$ or $\beta_0 = \gamma$. This expansion applies if $\gamma \neq 0$ and $\varepsilon \neq 0$. The series is right-hand side terminated for some non-negative integer $N$ if $P_N = 0$ and $c_{N+1} = 0$. If $\beta_0 = \alpha/\varepsilon$, the condition $P_N = 0$ is satisfied if

$$\alpha/\varepsilon = -N \quad \text{or} \quad \delta = -N. \quad (28)$$

If $\beta_0 = \gamma$, the only choice, since $\gamma$ should not be a negative integer number, is

$$\gamma + \delta - \alpha/\varepsilon = -N. \quad (29)$$

Another useful expansion is written in terms of the Tricomi confluent hypergeometric functions [39]:

$$H_C(\gamma,\delta,\varepsilon;\alpha,q;z) = \sum_{n=0}^{\infty} c_n U(\alpha/\varepsilon;\gamma-n;-\varepsilon z), \quad (30)$$

where the coefficients $c_n$ are given by the three-term recurrence relation (24) with

$$R_n = -\varepsilon n, \quad (31)$$

$$Q_n = (\delta+n)(1+n-\gamma) + \varepsilon n + \alpha - q, \quad (32)$$

$$P_n = (\delta+n)(\gamma-n-1-\alpha/\varepsilon). \quad (33)$$

This expansion applies if $\alpha \neq 0$ and $\varepsilon \neq 0$. The series may right-hand side terminate at some non-negative integer $N$ if $P_N = 0$, i.e.,

$$\gamma - 1 - \alpha/\varepsilon = N \quad \text{or} \quad \delta = -N. \quad (34)$$

For termination of above series the accessory parameter $q$ should necessarily fulfill the equation $c_{N+1} = 0$. This equation is convenient to write in the following matrix form:

$$\begin{bmatrix} Q_0 & R_1 & 0 & & \\ P_0 & Q_1 & R_2 & 0 & \\ 0 & P_1 & Q_2 & R_3 & \\ & \ddots & \ddots & \ddots & \\ & & 0 & P_{N-1} & Q_N \end{bmatrix} \begin{bmatrix} a_0 \\ a_1 \\ a_2 \\ \vdots \\ a_N \end{bmatrix} = \begin{bmatrix} 0 \\ 0 \\ 0 \\ \vdots \\ 0 \end{bmatrix}. \qquad (35)$$

The vanishing of the determinant of the above matrix gives a polynomial equation of the order $N+1$ for the accessory parameter $q$ for which the termination occurs.

### 4. Constant detuning models: real $z(t)$

Many specific subfamilies can be generated by the appropriate choice of $z(t)$. Consider first the case of constant detuning families of pulses generated by *real* functions $z(t)$. The families of pulses corresponding to $\delta_t(t) = \Delta = \text{const}$ are defined parametrically as:

$$t - t_0 = \frac{\delta_0}{\Delta} z + \ln z^{\delta_1/\Delta} + \ln(1-z)^{\delta_2/\Delta}, \qquad (36)$$

$$U(t) = \Delta \frac{U_0^* z^{k_1+1} (z-1)^{k_2+1}}{\delta_0 z^2 + (-\delta_0 + \delta_1 + \delta_2) z - \delta_1}. \qquad (37)$$

With an appropriate choice of parameters, Eq. (36) defines one-to-one mapping of the axis $t$ onto the interval $z \in (0,1)$. We specify the integration constant $t_0$, which actually produces only a shift in time, demanding $z(t=0) = 1/2$, hence, $t_0 = ((\delta_1 + \delta_2)\ln 2 - \delta_0/2)/\Delta$.

The derived families of pulses include both symmetric and asymmetric members. The amplitude modulation functions may or may not vanish at infinity. There are only six families for which the pulses vanish so that the pulse area is finite. These are the families with $k_{1,2} \neq -1$ which present, in general, asymmetric one- or two-peak pulses of controllable width. We will see that the asymmetry and the peak heights are mostly defined by the parameters $\delta_{1,2}$, while the pulse width is mainly controlled by $\delta_0$. The transformations $z(t)$ and corresponding pulse shapes $U(t)$ for the classes $k_{1,2} = 0$ and $k_{1,2} = -1/2$ at different values of parameters $\delta_{0,1,2}$ (in units of $\Delta$) are shown in Fig.2.

The family $k_{1,2} = -1/2$ represents generalization of the known family of Bambini and Berman [9] which corresponds to the choice $\delta_0 = 0$ (curves 1 in Fig.2, **b1**,**b2**,**b3**). In order to get an initial insight into how essential the addition of the $\delta_0$ term is we compare the graphs in Fig.2, **a1**,**a2**,**a3** and note the following: (i) the greater $\delta_0$ is, the wider the pulse, (ii) the smaller the parameters $\delta_1$ and $\delta_2$ are, the closer the pulse shape is to a rectangular form.

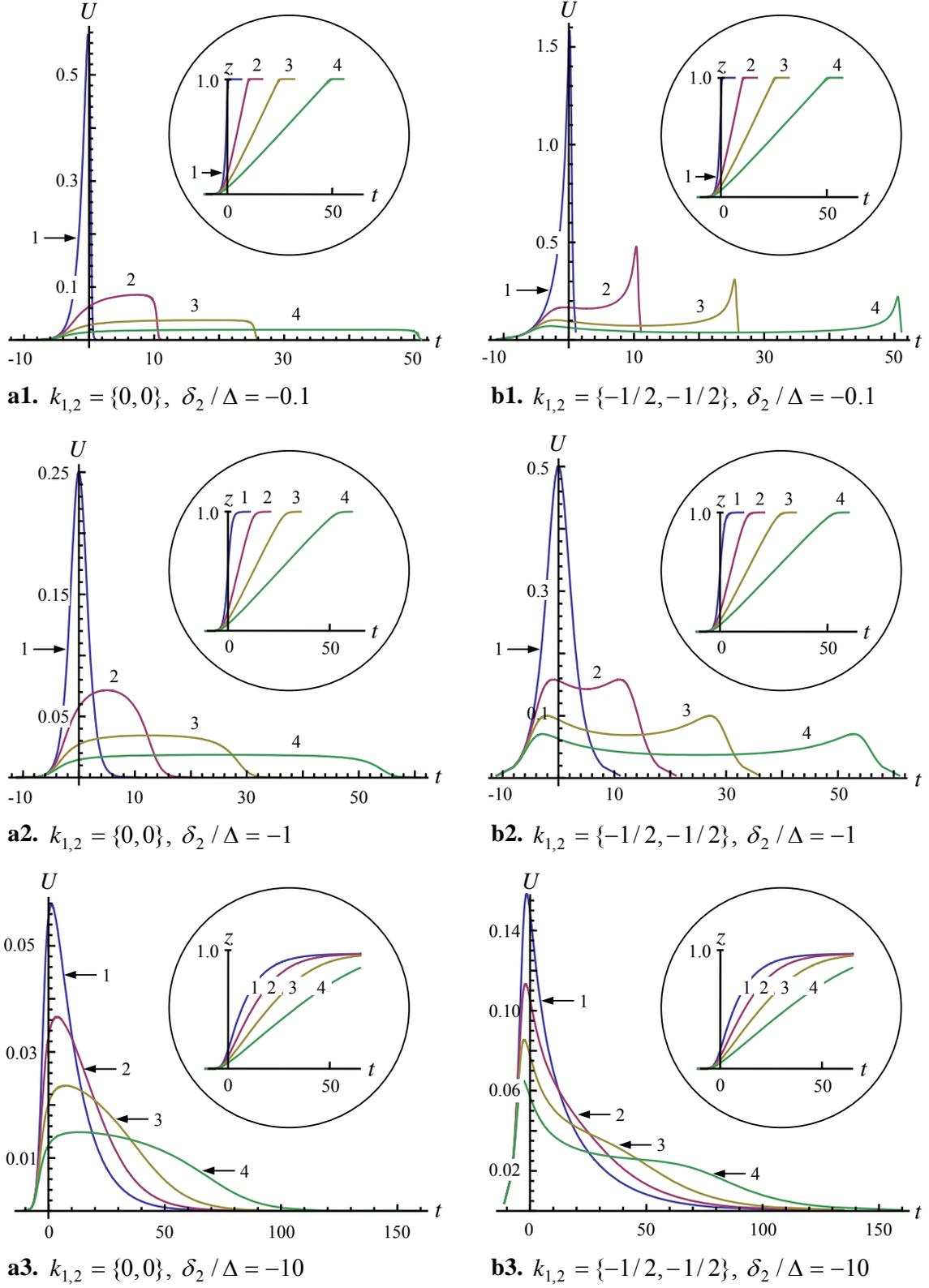

Fig. 2. Constant-detuning case $\delta_t = $ const: pulse shapes $U(t)$ and corresponding transformations $z(t)$ for the classes $k_{1,2} = 0$ (**a1-a3**) and $k_{1,2} = -1/2$ (**b1-b3**). $\delta_1/\Delta = 1$ and $\delta_0/\Delta = 0; 10; 25; 50$ (curves 1,2,3,4, respectively).

To make more explicit this observation, one-parametric subfamilies of symmetric-pulses belonging to the class $k_{1,2} = 0$ are shown in Fig.3, **a**,**b**. Here the parameters $\delta_{1,2}$ are fixed as $\delta_1 = -\delta_2$ and the subfamilies are parameterized only by $\delta_0$. The pulses are normalized to the same level and aligned horizontally to a common center. As we can see, these are smooth bell-shaped pulses (Fig.3, **a**) with different widths corresponding to different values of $\delta_0$. As $\delta_{1,2}$ approach zero, the bell shape becomes more rectangular (Fig.3, **b**), making it a better approximation for a rectangular box pulse. For the simultaneous limit $\delta_{1,2} \to 0$, the pulse becomes a step-wise function of time; that is, exact rectangular profile is achieved (which is, however, non-analytic itself at the edges).

Obviously, the pulse diverges if the denominator $P(z) = \delta_0 z^2 + (-\delta_0 + \delta_1 + \delta_2)z - \delta_1$ in the right-hand side of Eq. (37) vanishes at some $z_0$ on the interval $z \in (0,1)$. Then, after being normalized to $U_{\max} = 1$, it becomes infinitely narrow (Fig. 3a, curve 5). With one-to-one mapping $t \leftrightarrow z$ infinitely narrow pulse is possible only if $z_0$ is a multiple root of $P(z)$.

Consider the behavior of the pulse edges at $\delta_{1,2} \to 0$ in detail. In the limit $z \to 0$ the first and third terms in Eq. (36) are small compared with the second one. Neglecting these terms, however, gives the transformation $z(t) = e^{\Delta(t-t_0)/\delta_1}$ which leads to a diverging pulse.

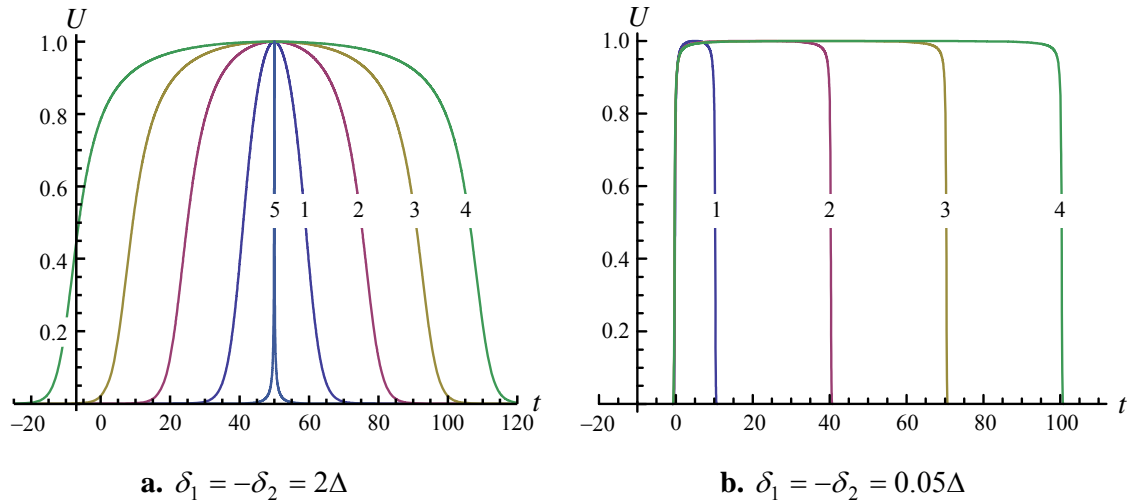

**a.** $\delta_1 = -\delta_2 = 2\Delta$   **b.** $\delta_1 = -\delta_2 = 0.05\Delta$

Fig. 3. Constant-detuning case $\delta_t = \Delta$, real $z(t)$. Pulse shapes $U(t)$ for the class $k_{1,2} = \{0,0\}$. $\delta_0 / \Delta = 10;\ 40;\ 70;\ 100$ (curves 1,2,3,4, respectively). The pulse width diverges as $\delta_0 \to \infty$, and infinitely narrow pulse is achieved when $\delta_0 = -4\delta_1$ (curve 5).

To get better approximation for small $z \ll 1$, one may expand $\ln(1-z)$ in Eq. (36) in power series. Then, keeping only the first term of the expansion we have

$$t - t_0 = ((\delta_0 - \delta_2)z + \delta_1 \ln z)/\Delta, \qquad (38)$$

which gives the transformation

$$z(t) = \frac{\delta_1}{\delta_0 - \delta_2} W\left(\frac{\delta_0 - \delta_2}{\delta_1} e^{\Delta(t-t_0)/\delta_1}\right), \qquad (39)$$

where $W$ is the Lambert $W$-function also known as product logarithm [48]. The pulse shapes generated by this function are compared with the exact ones defined by Eq. (36) in Fig. 4. We see that the two pulses are almost indistinguishable in the vicinity of the left edge for any allowed set of the involved parameters. Taking the limit $\delta_1 \to 0$ we see that the left edge becomes step-wise with a vertical jump located at $t_l = t_0|_{\delta_1 \to 0}$ or $t_l = (\delta_2 \ln 2 - \delta_0/2)/\Delta$. Similarly, in the limit $\delta_2 \to 0$ the right edge becomes step-wise with a vertical jump located this time at $t_r = t_0|_{\delta_2 \to 0} + \delta_0/\Delta$ or $t_r = (\delta_1 \ln 2 + \delta_0/2)/\Delta$. Hence, in the simultaneous limit $\delta_{1,2} \to 0$ the pulse width is $t_r - t_l = \delta_0/\Delta$. This limiting value for the pulse width can be obtained using the limiting exponential transformation mentioned above.

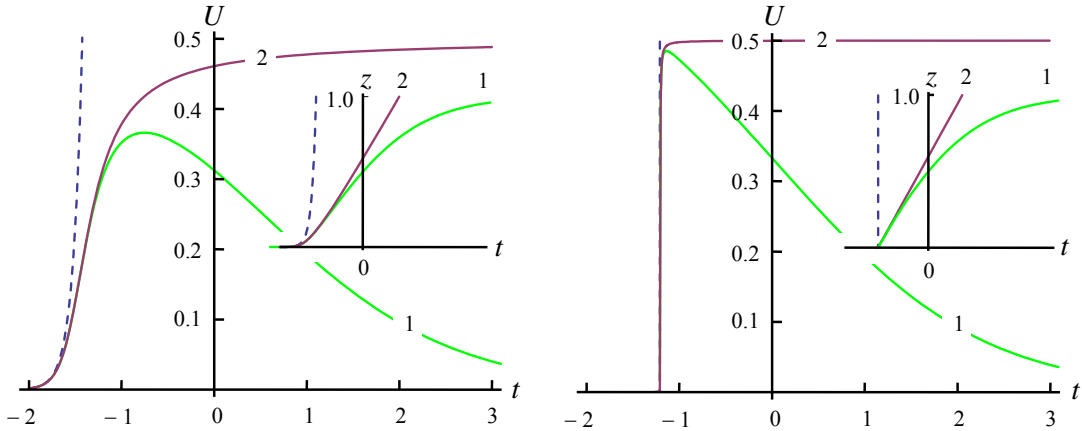

Fig. 4. Pulse shapes $U(t)$ and transformations $z(t)$ corresponding to Eqs. (36) and (39) (curves 1,2, respectively). The dashed lines represent the limiting exponential transformation $z(t) = e^{\Delta(t-t_0)/\delta_1}$. $U_0 = 1$, $\delta_0 = -\delta_2 = \Delta$. $\delta_1 = 0.1\Delta$ on the left and $\delta_1 = 0.001\Delta$ on the right.

## 5. Constant detuning models: complex-valued $z(t)$

A different set of constant-detuning subfamilies of pulses is generated by the *complex-valued* transformation $z = (1 + i y(t))/2$. With this transformation, real amplitude-modulation functions are generated only in three cases, when $k_1 = k_2$. This time, the pulse shapes are given parametrically as

$$t = \lambda_0 y + \lambda_1 \ln(1 + y^2) + 2\lambda_2 \arctan(y), \qquad (40)$$

$$U(t) = \frac{U_0 (1 + y^2)^{k_1+1}}{\lambda_0 + 2\lambda_2 + 2\lambda_1 y + \lambda_0 y^2}, \quad k_{1,2} = -1, -1/2, 0, \qquad (41)$$

where we have supposed $y(0) = 0$ and introduced new real parameters $\lambda_{0,1,2}$ and $U_0$: $\delta_0/\Delta = -2i\lambda_0$, $\delta_{1,2}/\Delta = \lambda_1 \mp i\lambda_2$, $U_0^* = -(2i)^{1+2k_1} U_0$. With an appropriate choice of parameters, Eq. (40) defines one-to-one mapping of the $t$-axis to the axis $y \in (-\infty, +\infty)$. Then, Eq. (41) defines asymmetric pulses, shown in Fig. 5. Note that the pulses of the subfamily $k_{1,2} = 0$ do not vanish at $t \to \pm\infty$: $U(\pm\infty) = U_0/\lambda_0$, while the subfamilies $k_{1,2} = -1/2$ and $k_{1,2} = -1$ present bell-shaped asymmetric pulses vanishing at infinity.

Though the qualitative behavior of the pulses in the last two cases is rather similar to those discussed by Bambini and Berman [9], for theoretical considerations the presented families may be more convenient because here the parameters of the confluent Heun function may be real so that in some cases closed form solutions can be derived using series expansions. A representative example for this observation is the case of the excitation of a two-level atom by a Lorentzian pulse (class $k_{1,2} = -1$, $\lambda_0 = 1$, $\lambda_{1,2} = 0$):

$$U(t) = \frac{U_0}{1 + t^2}, \quad \delta_t(t) = \Delta_0 = \text{const}. \qquad (42)$$

In this case $y = t$, $\delta_0 = -2i\Delta_0$, $\delta_{1,2} = 0$, $U_0^* = iU_0/2$, and the general solution of the problem can be written as

$$a_2 = (z/(z-1))^{U_0/2} \begin{bmatrix} C_1 H_C(1+U_0, 1-U_0, -2\Delta_0; 0, -U_0\Delta_0; z) + \\ C_2 e^{2\Delta_0 z} H_C(1+U_0, 1-U_0, 2\Delta_0; 4\Delta_0, (2+U_0)\Delta_0; z) \end{bmatrix}, \qquad (43)$$

where $z = (1+it)/2$. Since the parameters of the involved confluent Heun functions are real, we may apply the above series expansions in terms of the Kummer and Tricomi confluent hypergeometric functions to derive finite sum solutions. Note that here $\delta = 1 - U_0$ so that if the Rabi frequency $U_0$ is an integer number, the series may terminate for certain values of the detuning $\Delta_0$. The cases $U_0 = 1$ and $U_0 = 2$ produce the trivial result $\Delta_0 = 0$ (exact resonance), however, starting from $U_0 = 3$, the termination conditions lead to useful closed form exact solutions. The termination of the series is achieved if $\Delta_0 = 0, \pm 2\sqrt{3}$ for $U_0 = 3$, $\Delta_0 = 0, \pm 3/\sqrt{2}$ for $U_0 = 4$, etc; the number of non-zero terminating values of $\Delta_0$ is $(U_0 - 1)$ for odd $U_0$ and $(U_0 - 2)$ for even $U_0$. The solution of the two-state problem in these cases is written in terms of elementary functions. For instance, the result for $U_0 = 3$ is

$$a_2 = C_1 \frac{3t + 2t^2 + i\Delta_0/2}{2(1+t^2)^{3/2}} + C_2 \frac{e^{i\Delta_0 t}(1 + i\Delta_0 t/2)}{2(1+t^2)^{3/2}}, \quad \Delta_0 = \pm 2\sqrt{3}. \tag{44}$$

Note that here $C_1$ and $C_2$ are arbitrary constants, so that this is the general solution of the problem applicable for any initial condition. If the initial conditions $a_1(-\infty) = 1$, $a_2(-\infty) = 0$ are considered, $C_1$ becomes zero and only the second term remains in Eq. (44). Interestingly, it turns out that for this solution $a_2(+\infty) = 0$, hence, the parameter set $\{U_0, \Delta_0\} = \{3, \pm 2\sqrt{3}\}$ defines one of the complete return resonances when the system returns to its initial state at the end of the interaction. In the case of the Rabi model [4] the complete return spectrum is a periodic function of $U_0$ for any fixed $\Delta_0$. The same feature is also observed for the Rosen-Zener model [5]. Bambini and Berman have shown that return resonances in general do not occur for asymmetric pulses [9]; however, it was expected that the periodicity should be a feature of the spectrum whenever it exists, at least, for symmetric pulses. However, the case of the Lorentzian pulse clearly violates this supposition. This is readily verified using the exact finite sum solutions derived from the expansion of the confluent Heun function in terms of the Tricomi irregular confluent hypergeometric functions.

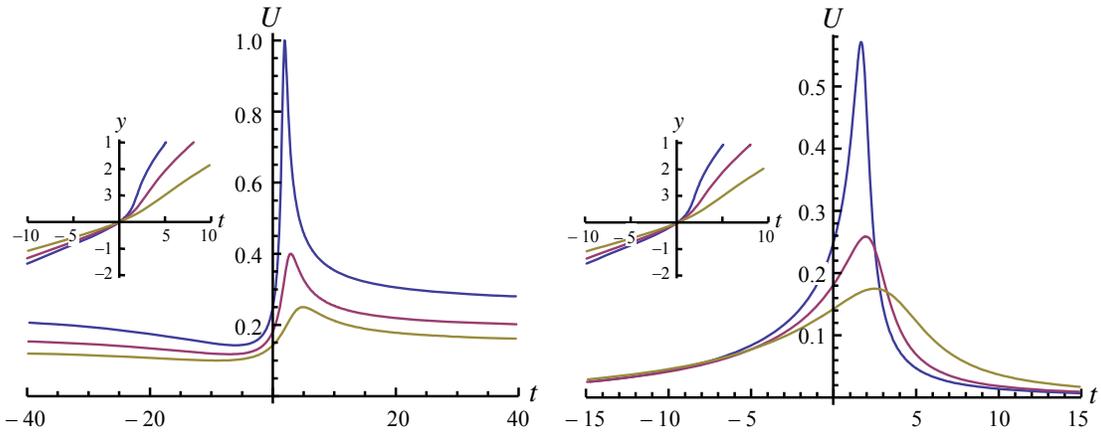

Fig. 5. Constant-detuning case: $\delta_t = \Delta$, complex transformation $z(t) = (1 - iy(t))/2$. Pulse shapes $U(t)$ for the classes $k_{1,2} = 0$ (on left) and $k_{1,2} = -1$ (on right) for $\lambda_1/\Delta = -3$, $\lambda_2/\Delta = 0$ and $\lambda_0/\Delta = 4;5;7$ (curves 1,2,3, respectively).

## 6. Resonance-crossing field configurations

Analyzing the derived classes in the general case of variable amplitude- and detuning-modulation functions we note that the most notable feature provided by the extension of the solvable models to the cases covered by the confluent Heun functions is due to the extension of the detuning modulation functions, which is mathematically manifested by the additional term $\delta_0$ in Eq. (12). Though simple at first glance, this term provides a significantly larger range of physically interesting detuning functions as compared with the cases when the problem is solved in terms of hypergeometric functions. We have already been convinced in that when discussing the constant detuning case $\delta_t(t) = \Delta$. We now will see that by different choices of the independent variable transformation $z = z(t)$ (both real and complex) many other interesting field configurations are modeled by the derived classes with $\delta_t(t) \neq \text{const}$, including multiple crossings of the resonance.

First, a specific aspect of this extension is that due to the $\delta_0$-term the detuning modulation function now allows resonance-crossings at two time-points. This can be easily understood by rewriting $\delta_z^*$ in the following form:

$$\delta_z^* = \frac{\delta_0 z^2 + (-\delta_0 + \delta_1 + \delta_2)z - \delta_1}{z(z-1)}. \tag{45}$$

Now, it is seen that the laser frequency detuning function turns into zero at the points

$$z_{1,2} = \frac{(\delta_0 - \delta_1 - \delta_2) \pm \sqrt{(-\delta_0 + \delta_1 + \delta_2)^2 + 4\delta_0 \delta_1}}{2\delta_0} \tag{46}$$

if the discriminant $D = (-\delta_0 + \delta_1 + \delta_2)^2 + 4\delta_0 \delta_1 \geq 0$. Hence, if $z_{1,2}$ are inner points of the allowed variation range of $z = z(t)$ and $z'(t) \neq 0$ everywhere within this range, Eq. (12) defines a detuning with *two* crossings of the resonance if $z_1 \neq z_2$ and a configuration *touching* the resonance if $z_1 = z_2$. Otherwise, we have a *non-crossing* model if none of the roots (46) belongs to the variation range of $z(t)$ and a *chirped* pulse configuration if only one root is an inner point of the variation range of $z(t)$. Alternatively, a chirped field configuration is obtained if the numerator in Eq. (45) is canceled to a linear function. This happens if $\delta_0 = 0$ or $\delta_1 = 0$ or $\delta_2 = 0$. Particularly, it is the case for all the three-parametric hypergeometric or confluent hypergeometric models. Since the non-crossing and chirped models have been intensively studied in the past and are well presented in literature (see, e.g., [1-10]) we do not discuss those cases here. Instead, we present some examples of field configurations with two resonance-crossings and a specific model describing multiple (periodically repeated) crossings.

An example of a field configuration with up to two crossings is the one discussed in our recent paper [20]. This is the model, referred to as the generalized Rosen-Zener model since it includes the original Rosen-Zener model [5] as a particular constant-detuning case, given by the following field configuration:

$$U(t) = U_0 \text{sech}(t), \quad \delta_t(t) = \Delta_0 + \Delta_1 (\text{sech}(t))^2. \tag{47}$$

This model is a member of the class $k_{1,2} = -1/2$ ($U^*/U_0^* = 1/\sqrt{z(z-1)}$) obtained by the real transformation $z = (1+\tanh(t/\tau))/2$ and the specifications $\delta_0 = 2\Delta_1 \tau$, $\delta_1 = -\delta_2 = \Delta_0 \tau/2$, $U_0^* = iU_0\tau$, $\tau = 1$. As can be seen, the crossings are due to the parameter $\delta_0$. If $\Delta_1 = -\Delta_0$, the detuning does not actually cross the resonance, but touches it at $t = 0$, thus representing the case of *level-glancing* field configuration [19]. In the general case of two crossings, the detuning mimics the parabolic crossing model [49,50], however, suggesting a finite variation of the detuning instead of the diverging one implied by the exact parabolic model. Note that with the same specifications for $z(t)$ and $\delta_{0,1,2}$, hence, with the same detuning (47), the basic model $k_{1,2} = 0$ produces amplitude modulation $U(t) = U_0(\text{sech}(t))^2$, and in the case of $k_{1,2} = -1$ we have a constant Rabi frequency case: $U(t) = U_0$.

In general, the detuning functions are asymmetric. For instance, this is the case for the transformation $z = (1+\tanh(t/\tau))/2$ if $\delta_1 \neq -\delta_2$ (for simplicity, we put $\tau = 1$):

$$\delta_t(t) = (\delta_1 - \delta_2) - (\delta_1 + \delta_2)\tanh(t) + \delta_0(\text{sech}(t))^2/2. \tag{48}$$

In this case the parameters $\delta_1$ and $\delta_2$ define the asymptotes of the detuning at $t \to \mp\infty$ that are now not equal (see Fig. 6a).

Moreover, the detuning function may be asymmetric even if $\delta_t(-\infty) = \delta_t(+\infty)$ when an asymmetric transformation $z(t)$ is applied. Such a situation arises, for example, when the constant amplitude field configuration is considered for the classes for which the function $U^*(z)$ is not symmetric with respect to interchange $k_1 \leftrightarrow k_2$. This is shown in Fig. 6b for the class $k_{1,2} = \{-1,+1\}$. Here, the transformation $z = -W(e^{1-t/\tau})$ (for simplicity, we put $\tau = 1$) with $W$ being the Lambert function, produces

$$U(t) = U_0, \quad \delta_t(t) = \delta_0 - \frac{\delta_0 + \delta_1 + \delta_2}{1+W(e^{1-t})} + \frac{\delta_2}{\left(1+W(e^{1-t})\right)^2}. \tag{49}$$

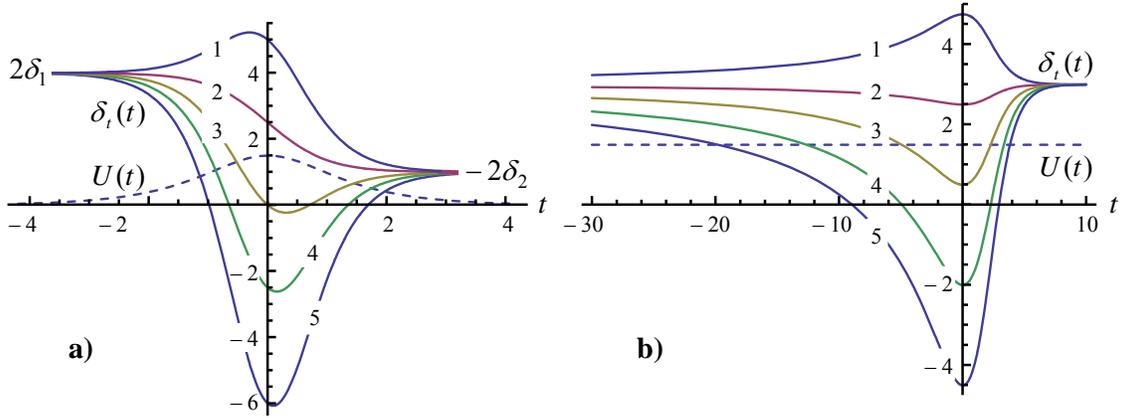

Fig. 6. Asymmetric detuning. All the parameters are assumed dimensionless.
(a) $k_{1,2} = \{-1/2; -1/2\}$, $U = U_0 \text{sech}(t)$, $U_0 = 1.5$, detuning is given by Eq. (48) with $\delta_1 = 2$, $\delta_2 = -1/2$, $\delta_0 = \{5; 0; -5; -10; -17\}$ (curves 1,2,3,4,5, respectively). (b) $k_{1,2} = \{-1, +1\}$, constant Rabi frequency case: $U = U_0 = 1.5$, detuning is given by Eq. (49) with $\delta_0 = 3$, $\delta_1 = -3$, $\delta_2 = \{-7; 2; 10; 20; 30\}$ (curves 1,2,3,4,5, respectively).

Finally, we complete this section by presenting an example of a constant amplitude model with periodically repeated resonance-crossings:

$$U(t) = U_0, \quad \delta_t(t) = A_0 \cos(\Delta t), \quad z(t) = (1 + \sin(\Delta t))/2. \quad (50)$$

This is a member of the class $k_{1,2} = -1/2$ with the parameters chosen as $U_0^* = iU_0/\Delta$, $\delta_0 = 2A_0/\Delta$, $\delta_{1,2} = 0$. Note that the amplitude of the detuning modulation, that is the maximum deviation $A_0$ of the detuning from the resonance $\delta_t = 0$, is controlled solely by $\delta_0$. This shows once more the usefulness of this parameter.

## 7. Generalized level-crossing Lorentzian model

Models with two crossings of the resonance are generated also by the complex-valued transformation $z = (1 + i y(t))/2$. An example due to the simplest choice $y(t) = t$ is the model referred to as the generalized Lorentzian model which describes the excitation of a two-level atom by a pulse of Lorentzian shape (compare with Eq. (42)):

$$U(t) = \frac{U_0}{1+t^2}, \quad \delta_t(t) = \Delta_0 + \frac{\Delta_1}{1+t^2}. \quad (51)$$

This configuration is generated by the basic model $k_{1,2} = -1$ with the specification $\delta_0 = -2i\Delta_0$, $\delta_1 = -\delta_2 = -i\Delta_1/2$ and $U_0^* = iU_0/2$. In this case the crossings are due to the parameters $\delta_{1,2}$. Note that if $k_{1,2} = -1/2$ is considered instead of $k_{1,2} = -1$, we will have $U(t) = U_0 / \sqrt{1+t^2}$, and $k_{1,2} = 0$ produces the constant Rabi frequency case $U(t) = U_0$. The time-variation of the detuning (51) for several values of the parameter $\Delta_1$, is shown in Fig. 7.

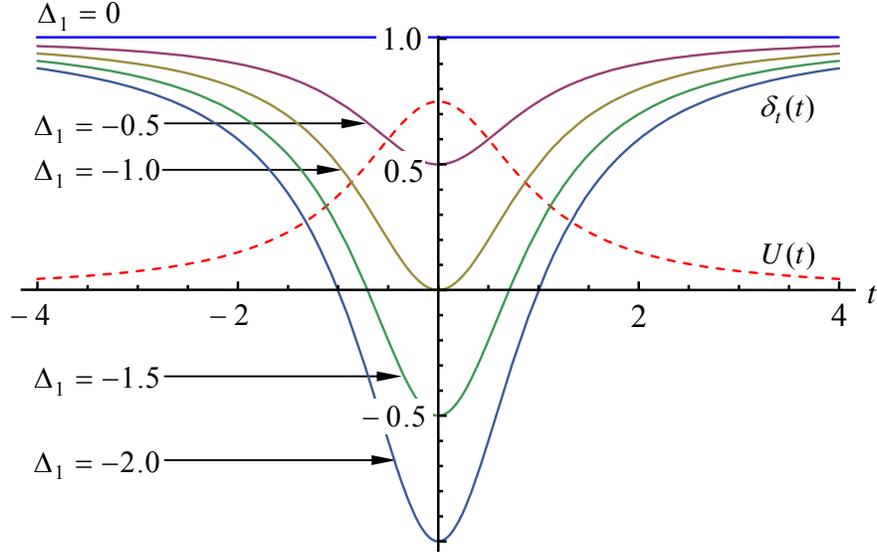

Fig. 7. Generalized Lorentzian model defined by Eqs. (51): $U_0 = 0.75$, $\Delta_0 = 1$. If $-\Delta_0 / \Delta_1 < 0$ or $-\Delta_0 / \Delta_1 > 1$, the detuning does not cross zero. If $\Delta_0 + \Delta_1 = 0$, it touches the origin. If $0 < -\Delta_0 / \Delta_1 < 1$, there are two crossing points located at $t = \pm\sqrt{-\Delta_1 / \Delta_0 - 1}$.

We have a family of symmetric detuning functions describing both non-crossing and crossing processes with one or two resonance crossing points. If $-\Delta_0 / \Delta_1 < 0$ or $-\Delta_0 / \Delta_1 > 1$, the detuning does not cross zero. If $\Delta_0 + \Delta_1 = 0$, the detuning touches the origin at $t = 0$ so that in this case we have a model of level-glancing [19]. If $0 < -\Delta_0 / \Delta_1 < 1$, there are two crossing points located at $t = \pm\sqrt{-\Delta_1 / \Delta_0 - 1}$. In this case the detuning again mimics the parabolic crossing model [49,50] as the generalized Rosen-Zener configuration (47). However, here also the variation of the detuning is within a finite region, not infinite as it is in the case of the exact parabolic model. Finally, note that $\Delta_0 = 0$ is a specific case when the detuning asymptotically goes to zero as $t \to \pm\infty$.

Though the above two symmetric detuning models, generalized Rosen-Zener and generalized Lorentzian, seem to offer similar double resonance-crossing field configurations, close inspection reveals that the very crossing processes have some qualitative differences. These are generated by the path that the variable $z$ draws on the complex plane. In the first case, $z$ goes from 0 to 1 along the real axis so that $z$ always belongs to the interval (0,1) connecting two regular singularities of the confluent Heun equation. In contrast to this, in the second case (the Lorentzian model), $z$ changes along the line passing through the midpoint of the interval $z \in (0,1)$ and goes parallel to the imaginary axis starting at $t = -\infty$ from irregular singular point $z = \infty$ of the Heun equation and returns back to the same irregular singularity at $t \to +\infty$. This is a rather complicated case since the behavior of the system is mostly governed by the irregular singularity, though the evolution of the system is strongly influenced by regular singularities $z = 0$ and $z = 1$.

However, fortunately, it turns out that in the case of the generalized Lorentzian model, again, as it was in the case of the constant detuning Lorentzian model, the solution of the problem involves a confluent Heun function with real parameters only. This allows one to get closed form solutions for some sets of the parameters involved using the above series solutions. Indeed, consider the general solution of the initial two-state problem. This solution can be written as a linear combination of two solutions of Eq. (2) constructed by using any two different sets of $\alpha_{0,1,2}$. For instance, taking $\alpha_0 = 0$ and $\alpha_0 = i\delta_0$ (see the first Eq. (17)), we have

$$a_2 = \left(\frac{z}{z-1}\right)^{\alpha_1} \big[ C_1 \cdot H_C(1+R, 1-R, -2\Delta_0; 0, -(R+\Delta_1/2)\Delta_0; z) \\ + C_2 e^{2\Delta_0 z} H_C(1+R, 1-R, 2\Delta_0; 4\Delta_0, (2+R-\Delta_1/2)\Delta_0; z) \big], \quad (52)$$

where $z = (1+it)/2$, $\alpha_1 = (\Delta_1 + 2R)/4$, $R = \sqrt{U_0^2 + \Delta_1^2/4}$ is the effective Rabi frequency, and $C_{1,2}$ are arbitrary constants. Accordingly, the series (23) and (30) may now terminate if the effective Rabi frequency is a natural number (since in that case $\delta = 1 - R = 0, -1, -2, ...$). The second termination condition then defines a relation between $\Delta_0$ and $\Delta_1$, for which the termination occurs:

$$\begin{aligned} R &= 1: \quad \Delta_0 = 0, \\ R &= 2: \quad \Delta_0 = \{0, -4/\Delta_1\}, \\ R &= 3: \quad \Delta_0 = \left\{0, -6/(\Delta_1 \pm \sqrt{3+\Delta_1^2/4})\right\}, ... \end{aligned} \quad (53)$$

Applying the expansions (23) and (30) to the first and the second terms in the right hand side of Eq. (52), respectively, we get that for the lowest order termination (i.e., when $\{R, \Delta_0\} = \{1, 0\}$) the general solution of the two-state problem (1) is explicitly written as

$$a_2 = e^{i(1+\Delta_1/2)\arctan(t)}\left(C_1 + \frac{C_2}{1+it}\right), \qquad (54)$$

and for the first non-trivial case $\{R,\Delta_0\} = \{2,-4/\Delta_1\}$ we have

$$a_2 = \left(\frac{1-it}{1+it}\right)^{-\Delta_1/4}\left(C_1\frac{8+4i\Delta_1 t + 8t^2 + \Delta_1^2}{1+t^2} + C_2\frac{e^{-4it/\Delta_1}}{1+t^2}\right). \qquad (55)$$

If the system starts from the first state, i.e., the initial condition $a_2(-\infty) = 0$ is considered, $C_1$ becomes zero and only the second term remains in these solutions. It is checked that then $a_2$ is zero also at $t = +\infty$, so that in this case the system returns to its initial state at the end of the interaction. The same feature, return of the system to its initial state at the end of the interaction if the system starts from the first level, is observed for all subsequent sets $\{R,\Delta_0\}$ given by Eqs. (53). Thus, the equations (53) define curves in the 3D space of the parameters $\{U_0,\Delta_0,\Delta_1\}$ belonging to the complete return spectrum of the system. The curves for $R = 2$ and $R = 3$ are shown in Fig. 8.

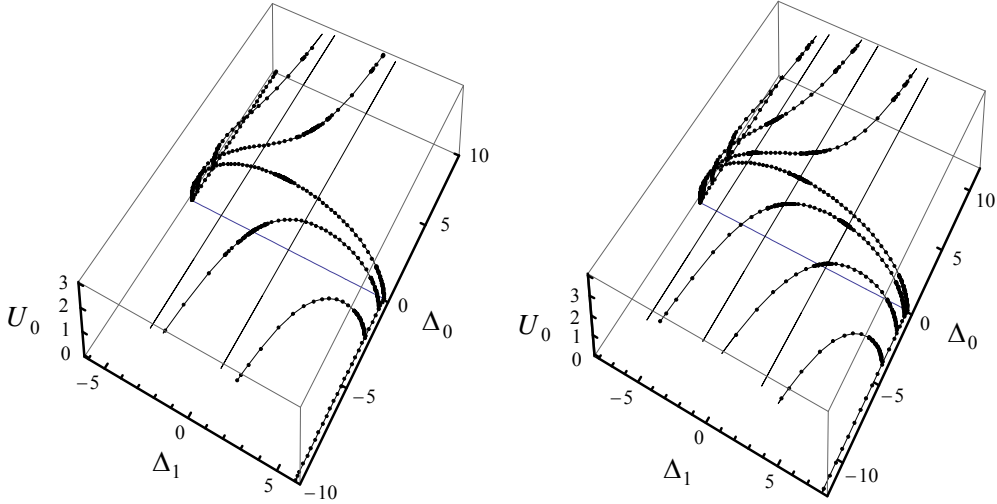

Fig. 8. Curves defined by Eqs. (53) for $R = 3$ and $R = 4$.

## 8. Summary

We have presented 15 classes of models allowing the solution of the time dependent two-state problem in terms of the confluent Heun functions. All the classes are *four-parametric* and include infinite number of members that are constructed by means of the application of an arbitrary chosen transformation of the dependent variable.

The obtained classes include all the known three- and two-parametric classes of two-state models solvable in terms of hypergeometric and confluent hypergeometric functions (note that these three- and two-parametric hypergeometric classes do not overlap). Nine of the classes extend the known three-parametric classes of models solvable in terms of the Gauss hypergeometric functions (there are six such classes) and in terms of the Kummer confluent hypergeometric functions (five classes) to a more general type of the detuning modulation function, including an additional parameter that allows the generation of a considerably wider variety of field configurations.

Discussing the *constant detuning* field configurations, we have shown that such models can be constructed by applying both real and complex transformations of the independent variable. In general, these are 4-parametric families of pulses including both symmetric and asymmetric members. The amplitude modulation functions may or may not vanish at infinity.

In the case of a *real* transformation, we identified six families for which the pulses vanish at infinity so that the pulse area is finite. We have shown that the asymmetry and the peak heights are mostly controlled by two of the involved parameters, $\delta_1$ and $\delta_2$, while the pulse width is mainly controlled by the third one, $\delta_0$. If $\delta_1$ goes to zero the left edge of the pulse becomes step-wise, and the same happens with the right edge if $\delta_2$ goes to zero. The simultaneous limit $\delta_{1,2} \to 0$ produces a box pulse. The pulse width diverges as $\delta_0 \to \infty$ and infinitely narrow pulse is achieved if $(-\delta_0 + \delta_1 + \delta_2)^2 - 4\delta_0\delta_1 = 0$.

A different set of constant-detuning subfamilies of pulses is generated by the *complex-valued* transformation of the independent variable. With this transformation, real amplitude-modulation functions are generated only in three cases, two of which provide bell-shaped asymmetric pulses vanishing at infinity. The qualitative behavior of these pulses is rather similar to those discussed by Bambini and Berman [9]; however, for theoretical considerations the presented families may be more convenient because here the parameters of the confluent Heun function may be real, allowing, in some cases, closed form solutions based on series expansions. We have presented an example supporting this observation. We have shown that in the case of the excitation of a two-level atom by a constant detuning laser pulse of Lorentzian shape, the solution of the corresponding two-state problem for some values of the involved parameters is written in terms of elementary functions.

Analyzing the derived classes in general, we note that the most notable features are due to an extra constant term in the detuning modulation function, e.g., numerous models with two resonance-crossing time points are due to this term. Double-crossing models are generated by both real and complex transformations of the independent variable. In general, the resulting detuning functions are asymmetric, the asymmetry being controlled by the parameters of the detuning modulation function. For some classes, however, the asymmetry may additionally be caused by the amplitude modulation function. We have presented an example of the latter possibility. Furthermore, we have mentioned a constant amplitude model with periodically repeated resonance-crossings.

Finally, we have discussed the excitation of a two-level atom by a pulse of Lorentzian shape. This model, referred to as the generalized Lorentzian model, suggests a family of symmetric detunings providing both non-crossing and crossing processes with one or two crossings of the resonance. We used particular series expansions of the solution of the confluent Heun equation in terms of the hypergeometric functions to derive particular closed form solutions of the two-state problem, both for the constant and variable detuning cases. The particular sets of the involved parameters for which these closed form solutions are obtained turned out to define curves in the 3D space of the involved parameters belonging to the complete return spectrum of the two-state quantum system.

There are very few papers discussing the solutions of the two-state problem in terms of the Heun functions. The biconfluent Heun equation was considered in [51] to generalize the models solvable in terms of the Kummer confluent hypergeometric functions and the general Heun equation was applied in [52] to study the two-state problem for an atom interacting with the bichromatic field of two lasers. As regards the confluent Heun function considered here, the problem was discussed, to the best of our knowledge, only in five papers. Three two-parametric families of pulses, for which, however, the involved confluent Heun functions are degenerated to the Kummer confluent hypergeometric or the Gauss hypergeometric functions are presented in [14] and [15], respectively. These are the subfamilies of the classes $k_{1,2} = \{-1/2, -1\}$, $\{1/2, -1\}$ and $\{(1/2, -1/2\}$, respectively. Other examples are the two three-parametric families discussed in [37,38] that belong to the classes $k_{1,2} = \{-1/2, 0\}$ and $\{0, -1/2\}$. In these cases, however, only the case $\delta_0 = 0$ was discussed. In the light of what has been revealed regarding the role of this parameter, this is, of course, a rather restrictive condition. Indeed, for instance, it is this parameter that controls the pulse width in the constant-detuning case, and it is this parameter that leads to double and periodic crossings of the resonance in the variable detuning case.

An additional methodological note is as follows. In deriving the presented classes we did not explicitly use the transformation of the independent variable. Rather, the stress was on the transformation of the dependent variable (see Eqs. (4),(6) and (7)). The transformation of the independent variable was used afterwards in order to generate particular families of pulses after the basic solutions of the integrable classes were identified. However, in most of the cases discussed in the literature only the transformation of the independent variable is applied. It should be said that this is a rather restrictive approach, which makes the treatment technically more complicated and, unfortunately, leads to a significantly narrower range of solvable cases. Indeed, examine the above 15 classes to see which one of them is possible to derive using only the independent variable transformation. In terms of the notations used above, it means that the pre-factor $\varphi(z)$ in the solution $a_2 = \varphi(z)u(z)$ is equal to unity, so that this is the case for which $\alpha_0 = \alpha_1 = \alpha_2 = 0$. This, in its turn, according to Eqs. (17), means that $k_1 \neq -1$, $k_2 \neq -1$ and $k_1 + k_2 \neq 0$. Only three classes out of the 15 derived ones meet these conditions, namely, the classes $k_{1,2} = \{-1/2, -1/2\}$, $\{-1/2, 0\}$, $\{0, -1/2\}$. These three cases are indicated by triangles in Fig.1. Besides, note that the case $k_{1,2} = \{-1/2, -1/2\}$ has a three-parametric subclass of models solvable in terms of the Gauss hypergeometric functions [9]. Note that, since this subclass already involves the parameters $U_0^*$, $\delta_1$ and $\delta_2$, in this case the only possible extension to produce new models not treated before may be due to a non-zero $\delta_0$. Thus, summarizing, we see that the above approach based on the transformation of the dependent variable provides significantly greater opportunities.

**Acknowledgments**

This research has been conducted within the scope of the International Associated Laboratory (CNRS-France and SCS-Armenia) IRMAS. The research has received funding from the European Union Seventh Framework Programme (FP7/2007-2013) under grant agreement No. 295025 – IPERA. The work has been supported by the Armenian State Committee of Science (SCS Grant No. 13RB-052).